\documentclass[pra,twocolumn]{revtex4}
\usepackage{graphicx}
\usepackage{amsfonts,amsmath,amssymb,float}
\usepackage{bm,dsfont}
\usepackage{color}
\usepackage{bbm}
\usepackage{longtable}
\usepackage[dvips]{epsfig}
\usepackage{amsmath,amssymb,lscape,float}
\usepackage{hyperref}
\usepackage{listings}

\def\tr{\qopname\relax{no}{Tr}}

\begin{document}
\title{Ground-state entanglement spectrum of a generic model with nonlocal 
excitation-phonon coupling}

\author{Vladimir M. Stojanovi\'c}
\email{vladimir.stojanovic@physik.tu-darmstadt.de} 
\affiliation{Institut f\"{u}r Angewandte Physik, Technical
University of Darmstadt, D-64289 Darmstadt, Germany}

\date{\today}

\begin{abstract} 
While the concept of the entanglement spectrum has heretofore been utilized to address various many-body systems,
the models describing an itinerant spinless-fermion excitation coupled to zero-dimensional bosons (e.g. dispersionless 
phonons) have as yet not received much attention in this regard. To fill this gap, the ground-state entanglement 
spectrum of a model that includes two of the most common types of short-ranged, nonlocal excitation-phonon 
interaction -- the Peierls- and breathing-mode couplings -- is numerically evaluated here. This model displays a 
sharp, level-crossing transition at a critical coupling strength, which signifies the change from a nondegenerate 
ground state at the quasimomentum $K_{\textrm{gs}}=0$ to a twofold-degenerate one corresponding to a symmetric 
pair of nonzero quasimomenta. Another peculiarity of this model is that in the special case of equal Peierls- and 
breathing-mode coupling strengths the bare-excitation Bloch state with the quasimomentum $0$ or $\pi$ is its exact 
eigenstate. Moreover, below a critical coupling strength this state is the ground state of the model. Thus, the sharp 
transition between a bare excitation and a heavily phonon-dressed (polaronic) one can be thought of as a transition 
between vanishing and finite entanglement. It is demonstrated here that the smallest ground-state entanglement-spectrum 
eigenvalue to a large extent mimics the behavior of the entanglement entropy itself and vanishes in this special case 
of the model; by contrast, all the remaining eigenvalues diverge in this case. The implications of excitation-phonon 
entanglement for $W$-state engineering in superconducting and neutral-atom-based qubit arrays serving as analog 
simulators of this model are also discussed.
\end{abstract}

\maketitle
\section{Introduction}
Over the past decade and a half, the concept of the {\em entanglement spectrum} established itself as a useful 
tool for understanding complex patterns of entanglement in strongly-interacting and/or topologically nontrivial quantum many-body 
systems~\cite{Laflorencie:16}. Starting from the pioneering work of Li and Haldane~\cite{Li+Haldane:08}, where this concept was 
utilized in the context of describing symmetry-protected topological states of matter~\cite{Fidkowski:10,Pollmann+:10,Thomale+:10}, 
low-lying entanglement spectra have been employed to glean nontrivial physical insights in many other areas of quantum
many-body physics. Examples include, but are not limited to, interacting spin chains~\cite{Poilblanc:10,Laeuchli+Schliemann:12,
Wu+:23,Yan+Meng:23}, integer quantum Hall effect~\cite{Schliemann:11}, interacting bosons~\cite{Deng+Santos:11,Ejima+:14} 
and fermions~\cite{Toldin+Assad:18}, many-body localization and thermalization in isolated quantum 
systems~\cite{Yang+:15,Geraedts+:17}, and Floquet dynamical phase transitions~\cite{Jafari+Akbari:21,Zhou:22,Luan+:2223}.

Early attempts to employ quantum information-theoretic tools (e.g. entanglement measures) in the description of 
many-body systems pertained to the use of entanglement entropy to characterize various quantum phase transitions~\cite{Amico:08}. 
The entanglement entropy~\cite{Wehrl:78,HayashiBOOK} of a quantum system that can be partitioned into two entangled 
subsystems (a bipartite quantum system) is given by the von Neumann entropy of the reduced density matrix corresponding 
to either one of the two subsystems (obtained by tracing out the other subsystem)~\cite{Laflorencie:16}. While this 
quantity does represent a measure of entanglement in any given state of such a system, the full entanglement 
spectrum -- i.e., the spectrum of the negative logarithm of the reduced density matrix~\cite{Chandran+:14} -- yields 
a much more detailed characterization of the entanglement in the system.

This paper is focussed on the ground-state entanglement spectrum of a one-dimensional (1D) lattice model describing 
an itinerant spinless-fermion excitation coupled to zero-dimensional bosons (e.g., dispersionless phonons) through 
two different short-ranged coupling mechanisms. The entanglement aspects of coupled  excitation-phonon (e-ph) 
models -- not least those pertaining to entanglement spectra -- have heretofore not been given due 
attention~\cite{Stojanovic:20,Roosz+Held:22}. While the dearth of entanglement-related studies of such models 
is the primary motivation behind the present work, the choice of e-ph interaction mechanisms to be discussed 
here -- namely, Peierls (P) and breathing-mode (BM) type couplings -- is motivated by their relevance in various 
physical systems. 

Models with short-ranged e-ph coupling describe the interaction of an excess charge carrier (or an 
exciton) in certain classes of electronic materials (e.g., narrow-band semiconductors) with optical phonons 
of the host crystal~\cite{Hannewald++:04,Hannewald:04,Roesch+:05,Slezak++:06,Vukmirovic+:10,Stojanovic+:10,
Vukmirovic+:12,Shneyder+:20,Shneyder+:21}. In the extreme case of strong e-ph coupling, a heavily phonon-dressed 
excitation is formed (small polaron)~\cite{RanningerReview:06,AlexandrovDevreeseBook}. While the bulk of 
small-polaron-related studies~\cite{Ranninger:92,Capone+:97,Wellein+Fehske,Jeckelmann+White:98,Bonca+:99,Zoli:03,
Stojanovic+:04,Stojanovic+:12,Mei+:13,Chakraborty+:16,Jansen+:19} have so far been carried out within the framework
of the Holstein model~\cite{Holstein:59}, which accounts for the purely local e-ph coupling, over the 
past two decades considerable attention was devoted to nonlocal-coupling mechanisms~\cite{Zoli:03,Stojanovic+:04}. 
The most well-known among them is P-type coupling~\cite{Stojanovic+:04,Stojanovic:08}, which is manifested through 
the dependence of the dynamically-fluctuating excitation hopping amplitude between two adjacent lattice sites on
the difference of Einstein-phonon displacements on those sites. On the other hand, BM-type coupling accounts for 
a change in the on-site energy of an itinerant excitation due to phonon displacements on the two adjacent sites; 
being also of the density-displacement type, it can be viewed as a nonlocal counterpart of Holstein-type coupling. 

What makes strongly momentum-dependent e-ph interactions -- those like P-type coupling, whose vertex function 
depends both on the excitation- and phonon quasimomenta -- particularly interesting is the fact that they
allow the possibility of sharp, level-crossing transitions at a critical e-ph coupling strength; according 
to the Gerlach-L\"{o}wen theorem~\cite{Gerlach+Lowen:87,GerlachLowenRMP:91}, transitions of this type are 
forbidden for momentum-independent (Holstein-type) coupling and those that depend on the phonon quasimomentum, 
but not on that of the excitation (e.g., Fr\"{o}hlich-type coupling~\cite{Froehlich:54}). Such a sharp transition 
corresponds to a change from a nondegenerate ground state at the quasimomentum $K_{\textrm{gs}}=0$ to 
a twofold-degenerate one corresponding to a pair of equal and opposite (nonzero) quasimomenta. It was already
demonstrated such transitions take place in a model with P-type coupling~\cite{Stojanovic:08}, as well 
as in its counterpart with simultaneous P and BM couplings discussed in the present work~\cite{Stojanovic+:14,Stojanovic+Salom:19}.

Aside from the occurrence of a sharp ground-state transition, another peculiar feature of the model under 
consideration -- with simultaneous P and BM couplings -- is that in the special case when the two coupling 
strengths are equal the bare-excitation Bloch state with the quasimomentum $k=0$ (or $k=\pi$) is its eigenstate. 
Furthermore, below a critical coupling strength this bare-excitation state -- with no e-ph entanglement -- represents
the ground state of the model. In the two existing proposals for the physical realization of this model -- with 
superconducting qubits coupled to resonators~\cite{Stojanovic+:14,Stojanovic+Salom:19,Nauth+Stojanovic:23} 
and neutral-atom-based~\cite{Stojanovic:21} Rydberg-dressed qubit arrays -- this feature translates into 
the possibility of engineering $W$-type states~\cite{Stojanovic:21,StojanovicPRL:20}, maximally-entangled 
multiqubit states of interest for quantum-technology applications. 

Here the ground-state entanglement spectrum of a model with simultaneous P and BM couplings is studied in a numerically-exact 
fashion, the primary aim of this study being to describe the dependence of this spectrum on the effective e-ph coupling 
strength. In addition to verifying that each entanglement-spectrum eigenvalue undergoes nonanalyticities at critical 
coupling strengths, it is demonstrated here that the entanglement entropy around this transition point is predominantly 
determined by the smallest eigenvalue. This eigenvalue to a large extent mimics the behavior of the entropy and vanishes 
in the special case of equal P and BM coupling strengths; at the same time, all the remaining entanglement-spectrum 
eigenvalues diverge for equal coupling strengths, while still yielding vanishing contributions to the entanglement 
entropy. The implications of the behavior of the entanglement spectrum and the corresponding entropy in the latter 
case for the generation of $W$ states in the previously proposed analog quantum simulators of the model under consideration 
are also discussed.
%It is also demonstrated that -- as a consequence of the discrete translational 
%symmetry of the system -- the eigenvalues from the entanglement spectrum can be labeled by the bare-excitation quasimomentum 
%quantum numbers. This is complemented by the numerical finding that this quantum number in the model under consideration takes 
%values $0$ and $\pi$, including cases where a transition between the two occurs at a coupling strength far larger 
%than the critical one. The entanglement spectrum in this special case of 
%the model is compared here with its counterpart in the generic case. The implications of excitation-phonon 
%entanglement for quantum-state engineering in superconducting- and neutral-atom-based analog simulators of 
%this model are also discussed.

This article is organized as follows. In Sec.~\ref{Model} the coupled e-ph Hamiltonian -- comprising P- and BM 
coupling terms -- is introduced, along with a discussion of the bare-excitation eigenstates in a special case of this 
model and a short description of possible physical realizations of this model. Section~\ref{entspectrum} starts with a 
short recapitulation of the definition 
and general properties of entanglement spectra, followed by the essential details of their application in the coupled e-ph 
system under consideration. Some mathematical consequences of the discrete translational symmetry of the system, as well as
the relevant details of the truncation of the total Hilbert space of the system, are discussed in Sec.~\ref{CompStrategy}.
The principal results of the paper are presented and discussed in Sec.~\ref{ResultsDiscuss}. To end with, the paper 
is summarized in Sec.~\ref{SumConcl}. In order not to interrupt the main flow of the paper, some mathematical derivations
are relegated to Appendices~\ref{derivelambda} and \ref{RedDensMatEl}.

\section{Model} \label{Model}
To set the stage for further discussion, the coupled e-ph Hamiltonian with Peierls- and breathing-mode 
coupling terms is introduced, both in the real-space (lattice) and momentum-space representations 
(Sec.~\ref{ModelHamiltonian}). An interesting special case of this Hamiltonian -- namely, the one 
with equal coupling strengths for the two relevant e-ph interaction mechanisms -- is then briefly discussed 
(Sec.~\ref{BareExcEigenstate}). For the sake of completeness, this is followed by a short description 
of two proposed physical realizations of the model under consideration (Sec.~\ref{PhysRealize}).

\subsection{Hamiltonian and its ground-state properties} \label{ModelHamiltonian}
The 1D lattice model under consideration describes a single spinless-fermion excitation interacting 
with dispersionless (Einstein-type) phonons through two different nonlocal e-ph coupling mechanisms. 
The Hamiltonian of this model is given by $H=H_0+H_{\textrm{e-ph}}$, where $H_{0}$ is the noninteracting
and $H_{\textrm{e-ph}}$ the interacting (e-ph) part. The noninteracting part $H_{0}$ consists of the excitation 
kinetic-energy term, with the corresponding hopping amplitude $t_{\textrm{e}}$, and free-phonon terms ($\hbar=1$):
\begin{equation}\label{H_0}
H_{0} = -t_{\textrm{e}}\sum_n (c^\dagger_{n+1}c_n + \mathrm{H.c.}) 
+\omega_{\textrm{ph}}\sum_n a^\dagger_n a_n\:.
\end{equation}
Here $c^\dagger_n$ ($c_n$) creates (annihilates) an excitation at site $n$ ($n=1,\ldots, N$) of the underlying 
1D lattice, while $a^\dagger_n$ ($a_n$) creates (annihilates) a zero-dimensional (Einstein) phonon with frequency 
$\omega_{\textrm{ph}}$ at the same site. 

The total e-ph coupling part $H_{\textrm{e-ph}}$ of the total Hamiltonian consists of the Peierls- (P)
and breathing-mode (BM) contributions. The P contribution~\cite{Stojanovic+:04,Stojanovic:08}
accounts for the linear dependence of the effective (dynamically dependent on the phonon degrees of freedom)
excitation-hopping amplitude between sites $n$ and $n+1$ on the difference of the respective local phonon 
displacements. In its most succinct form, it is given by
\begin{equation}\label{H_P}
H_{\textrm{P}} = g_{\textrm P}\omega_{\textrm{ph}}\:l_0^{-1}\:
\sum_{n}(c_{n}^{\dagger}c_{n+1}+\textrm{H.c.})\:(u_{n+1}-u_{n}) \:,
\end{equation}
where $g_{\textrm{P}}$ is the corresponding dimensionless P-coupling strength and 
$u_n \equiv l_0(a_{n}+a_{n}^{\dagger})$ the phonon displacement at site $n$, with 
$l_0$ being the phonon zero-point length. At the same time, the BM contribution 
captures the antisymmetric nonlocal coupling of the excitation density at site $n$ 
with the local phonon displacements on the nearest-neighbor sites $n\pm 1$~\cite{Slezak++:06,Stojanovic+:14}. 
This density-displacement type coupling term is given by
\begin{equation}\label{H_BM}
H_{\textrm{BM}} = g_{\textrm{BM}}\omega_{\textrm{ph}}\:l_0^{-1}\:
\sum_n c_{n}^{\dagger}c_{n}(u_{n-1}-u_{n+1}) \:,
\end{equation}
where $g_{\textrm{BM}}$ stands for the dimensionless BM-coupling strength.
In their most explicit forms, the coupling terms $H_{\textrm{P}}$ and $H_{\textrm{BM}}$ 
are given by [cf. Eqs.~\eqref{H_P} and \eqref{H_BM}]
\begin{eqnarray} \label{ExplicitFormH_PBM}
H_{\textrm{P}} &=& g_{\textrm{P}}\omega_{\textrm{ph}}\sum_n (c^\dagger_{n+1}c_n 
+ \mathrm{H.c.})  \nonumber \\
&\times& (a^\dagger_{n+1} + a_{n+1} - a^\dagger_n - a_n) \:, \\
H_{\textrm{BM}} &=& g_{\textrm{BM}}\omega_{\textrm{ph}}\sum_n c^\dagger_{n}c_n \nonumber \\
&\times& (a^\dagger_{n-1} + a_{n-1} - a^{\dagger}_{n+1} - a_{n+1})  \nonumber\:.
\end{eqnarray}

The total e-ph coupling Hamiltonian $H_{\textrm{e-ph}}=H_{\textrm{P}}+H_{\textrm{BM}}$ can be recast 
in the generic momentum-space form 
\begin{equation}\label{Heph_ms}
H_{\textrm{e-ph}}=\frac{1}{\sqrt{N}}\sum_{k,q}\gamma_{\textrm{e-ph}}(k,q)\:
c_{k+q}^{\dagger}c_{k}(a_{-q}^{\dagger}+a_{q})\:, 
\end{equation}
where the P- and BM contributions to the total e-ph vertex function 
$\gamma_{\textrm{e-ph}}(k,q)=\gamma_{\textrm{P}}(k,q) +\gamma_{\textrm{BM}}(q)$ 
are given by
\begin{eqnarray}\label{vertex_func}
\gamma_{\textrm{P}}(k,q) &=& 2ig_{\textrm{P}}\:\omega_{\textrm{ph}}\:[\sin k-\sin(k+q)] \label{gammaP} \:, \\
\gamma_{\textrm{BM}}(q)  &=& 2ig_{\textrm{BM}}\:\omega_{\textrm{ph}}\:\sin q \label{gammaBM} \:.
\end{eqnarray}
[Note that quasimomenta $k$ and $q$ in Eqs.~\eqref{gammaP} and \eqref{gammaBM} are assumed to be dimensionless, 
i.e., expressed in units of the inverse lattice period; this convention will be used throughout the remainder 
of this paper.]

For the most general ($k$- and $q$-dependent) vertex function $\gamma_{\textrm{e-ph}}(k,q)$, the 
effective coupling strength is given by 
\begin{equation} \label{genlambda}
\lambda_{\textrm{e-ph}}=\frac{\langle|\gamma_{\textrm{e-ph}}(k,q)|^{2}\rangle_{\textrm{BZ}}}
{2t_{\rm e}\:\omega_{\textrm{ph}}} \:, 
\end{equation}
where $\langle\ldots\rangle_{\textrm{BZ}}$ is the Brillouin-zone (BZ) average over the quasimomenta 
$k,q \in (-\pi,\pi]$. A straightforward derivation (for details, see Appendix~\ref{derivelambda}) 
leads to the following results for the effective P- and BM coupling strengths:
\begin{eqnarray}  \label{lambdaPandBM}
\lambda_{\textrm{P}}  =  2g^2_{\textrm{P}}\:\frac{\omega_{\textrm{ph}}}{t_{\rm e}} \qquad,\qquad
\lambda_{\textrm{BM}} =  g^2_{\textrm{BM}}\:\frac{\omega_{\textrm{ph}}}{t_{\rm e}}\:.
\end{eqnarray}

It is also important to notice that -- due to the specific momentum dependence of the two relevant 
(P and BM) e-ph couplings [cf. Eqs.~\eqref{gammaP} and \eqref{gammaBM}] -- the total effective 
e-ph coupling strength $\lambda_{\textrm{e-ph}}$ [cf. Eq.~\eqref{genlambda}] is given by the
sum of $\lambda_{\textrm{P}}$ and $\lambda_{\textrm{BM}}$, i.e. 
$\lambda_{\textrm{e-ph}}=\lambda_{\textrm{P}}+\lambda_{\textrm{BM}}$ 
(for details, see again Appendix~\ref{derivelambda}).

It is worthwhile pointing out that -- as a direct implication of the discrete translational symmetry 
of the system (regardless of the concrete form of $H_{\textrm{e-ph}}$) -- the eigenstates of the 
total Hamiltonian $H=H_0+H_{\textrm{e-ph}}$ ought to be good-quasimomentum states. More precisely, 
these states -- the Bloch eigenstates of the coupled e-ph system at hand -- are the joint eigenstates
of the total Hamiltonian $H$ and the total quasimomentum operator
\begin{equation}\label{totalcryst}
K_{\textrm{tot}}=\sum_{k} k\:c^{\dagger}_{k}c_{k}+\sum_{q}q\:a^{\dagger}_{q}a_{q} \:,
\end{equation}
because the latter commutes with $H$. In the following, the eigenvalues of $K_{\textrm{tot}}$ will be
labelled by $K \in (-\pi.\pi]$.  

The fact that the total e-ph vertex function in Eq.~\eqref{Heph_ms} depends on both the excitation ($k$) 
and phonon ($q$) quasimomenta implies that the Hamiltonian $H_{\textrm{e-ph}}$ does not satisfy the conditions 
for the applicability of the Gerlach-L\"{o}wen theorem, which rules out the existence of nonanalyticites in the 
ground-state-related quantities~\cite{GerlachLowenRMP:91}. In particular, the ground state of the total Hamiltonian 
$H=H_0+H_{\textrm{e-ph}}$ under consideration undergoes a sharp, level-crossing-type transition at a certain 
non-universal (i.e., dependent on the adiabaticity ratio $\omega_{\textrm{ph}}/t_{\rm e}$) critical 
value $\lambda^{\textrm{c}}_{\textrm{e-ph}}\sim 1$ of the effective coupling strength~\cite{Stojanovic:08}. 
Below this critical value (i.e. for $\lambda_{\textrm{e-ph}}<\lambda^{\textrm{c}}_{\textrm{e-ph}}$) the ground state 
is nondegenerate and represents the $K=0$ eigenvalue of $K_{\mathrm{tot}}$; on the other hand, for $\lambda_{\textrm{e-ph}}
\ge\lambda^{\textrm{c}}_{\textrm{e-ph}}$ the ground state is twofold-degenerate and corresponds to a pair of 
equal- and opposite (nonzero) quasimomenta. With $\lambda_{\textrm{e-ph}}$ increasing beyond its critical value 
(corresponding to the given value of the adiabaticity ratio), the pair of quasimomenta $K_{\textrm{gs}}$ that 
corresponds to the twofold-degenerate ground state also varies; this is reflected in the ground-state energy 
undergoing a sequence of further first-order nonanalyticities. Finally, this quasimomentum saturates at 
$K_{\textrm{gs}}=\pm\pi/2$ for $\lambda_{\textrm{e-ph}}$ above a threshold value; the latter also depends on 
$\omega_{\textrm{ph}}/t_{\rm e}$ and is larger than the corresponding $\lambda^{\textrm{c}}_{\textrm{e-ph}}$. 

\subsection{Bare-excitation Bloch eigenstates for $g_{\textrm{P}}=g_{\textrm{BM}}$} \label{BareExcEigenstate}
The model under consideration has an interesting property in the special case when the two 
relevant coupling strenghts are the same (i.e., $g_{\textrm{P}}=g_{\textrm{BM}}$). Namely, 
in this special case the coupled e-ph Hamiltonian of the system posseses a bare-excitation 
Bloch eigenstate for an arbitrary e-ph coupling strength. This eigenstate $|\Psi_{k}\rangle 
\equiv c^{\dagger}_{k}|0\rangle_{\textrm{e}}\otimes|0\rangle_{\textrm{ph}}$ (where $|0\rangle_{\textrm{e}}$ 
and $|0\rangle_{\textrm{ph}}$ are the excitation and phonon vacuum states, respectively) 
corresponds to the excitation quasimomentum $k=0$ in the case of positive hopping amplitude 
($t_e>0$), while for $t_e<0$ it corresponds to $k=\pi$. 

It is rather straightforward to show that $|\Psi_{k=0}\rangle\equiv c^{\dagger}_{k=0}
|0\rangle_{\textrm{e}}\otimes|0\rangle_{\textrm{ph}}$ is an eigenstate of the total Hamiltonian
for ($t_e>0$) (the proof that $|\Psi_{k=\pi}\rangle$ is its eigenstate for $t_e<0$ is completely 
analogous). Namely, because $|\Psi_{k=0}\rangle$ is an eigenstate of the noninteracting part $H_0$ 
of the system Hamiltonian (for $t_e>0$), to demonstrate that it is also an eigenstate of $H$ it is 
sufficient to prove that this state is simultaneously an eigenstate of $H_{\textrm{e-ph}}$ [cf. Eq.~\eqref{Heph_ms}]. 
By acting with $H_{\textrm{e-ph}}$ on $|\Psi_{k=0}\rangle$, taking into account that $c_{k}c_{k=0}^{\dagger}
\:|0\rangle_{\textrm{e}}\equiv\delta_{k,0}|0\rangle_{\textrm{e}}$, one readily finds that
\begin{equation}\label{vanish}
H_{\textrm{e-ph}}|\Psi_{k=0}\rangle = \frac{1}{\sqrt{N}}\sum_{q}\gamma_{\textrm{e-ph}}(k=0,q)\:
c_{q}^{\dagger}|0\rangle_{\textrm{e}}\otimes a_{-q}^{\dagger}|0\rangle_{\textrm{ph}}\:.
\end{equation}
Given that for the vertex functions in Eqs.~\eqref{Heph_ms}--\eqref{gammaBM} it holds that
(for an arbitrary phonon quasimomentum $q$)
\begin{equation}\label{VanCond}
\gamma_{\textrm{e-ph}}(k=0,q)\equiv\gamma_{\textrm{P}}(k=0,q)
+\gamma_{\textrm{BM}}(q)=0  \:,
\end{equation}
every term in the sum on the RHS of Eq.~\eqref{vanish} vanishes, 
which immediately implies that $H_{\textrm{e-ph}}|\Psi_{k=0}\rangle=0$. Thus, $|\Psi_{k=0}\rangle$ is an 
exact eigenstate of $H_{\textrm{e-ph}}$ for an arbitrary $\lambda_{\textrm{e-ph}}$ (its corresponding eigenvalue 
is equal to zero); this concludes the proof that $|\Psi_{k=0}\rangle$ is an eigenstate of $H$. 

While the state $|\Psi_{k=0}\rangle$ ($|\Psi_{k=\pi}\rangle$) is an eigenstate of $H$ in the $t_e>0$ ($t_e<0$)
case, it can be demonstrated numerically that it is also the ground state of this Hamiltonian below a certain critical 
coupling strength $\lambda^{\textrm{c}}_{\textrm{e-ph}}$ (for details, see Sec.~\ref{ResultsDiscuss} below). 
Because these bare-excitation states have no phonon content, these ground states correspond to the bare-excitation 
band minimum and the total quasimomentum in the cases is $K_{\textrm{gs}}=0$ and $K_{\textrm{gs}}=\pi$. The occurrence
of such, bare-excitation, ground state in a coupled e-ph model has two interesting implications. The first one is 
that -- unlike the more common (phonon-dressed) ground states of coupled e-ph models~\cite{Engelsberg+Schrieffer:63} -- this 
bare-excitation ground state is not accompanied by the usual one-phonon continuum of states starting from the energy $\hbar\omega_{\textrm{ph}}$ 
above the ground-state energy~\cite{Nauth+Stojanovic:23}. The second implication, of more direct interest for 
the present work, is that for $\lambda_{\textrm{e-ph}}<\lambda^{\textrm{c}}_{\textrm{e-ph}}$ the Hamiltonian $H$
has a separable ground state, i.e. a ground state with no entanglement between excitation- and phonon degrees of freedom. 

\subsection{Physical realizations with superconducting and neutral-atom qubits}  \label{PhysRealize}
In addition to its relevance for real electronic materials, the model under consideration can be 
realized with analog quantum simulators. Its realizations with an array of superconducting transmon 
qubits inductively coupled with microwave resonators~\cite{Stojanovic+:14,Stojanovic+Salom:19}, 
as well as with an array of cold neutral atoms in optical tweezers interacting through Rydberg-dressed 
resonant dipole-dipole interaction~\cite{Stojanovic:21}, have already been proposed.

In the transmon-based realization~\cite{Stojanovic+:14}, in which the role of phonons is 
played by microwave photons in resonators, the central idea is the use of a coupler circuit between 
each pair of qubits, which represents a generalization of a SQUID loop and mediates both qubit-qubit 
and qubit-resonator coupling in this system; while the qubit-qubit coupling turns out to be of the 
$XY$-type and maps -- via the Jordan-Wigner transformation -- into the hopping terms for an itinerant spinless-fermion
excitation, the qubit-resonator coupling (after the same transformation) gives rise to P- and BM coupling 
between this excitation and photons in the resonators; in particular, the term that is mapped into 
the P-coupling term in this way has the form characteristic of the $XY$ spin-Peierls model. The coupler 
circuit consists of two loops and three Josephson junctions, with both loops being subject to external 
magnetic fluxes. The upper loop of this circuit, delineated by two junctions with identical Josephson 
energies, is threaded by an external ac flux and the flux generated by the photon modes in resonators; 
it is the latter flux that gives rise to an inductive qubit-resonator coupling in this system. At the 
same time, the total flux in the lower loop of the coupler circuit -- whose two junctions have unequal Josephson 
energies -- consists of an external ac flux (with the same driving frequency but with a different 
magnitude and sign than its cunterpart in the upper loop) and a dc flux, the latter being the main 
externally tunable parameter (experimental knob) in the system. This realization yields -- by 
design -- identical P and BM coupling strengths [see Sec.~\ref{BareExcEigenstate} above].

The alternative realization of the model, with neutral atoms (e.g. Rb$^{87}$) in optical tweezers~\cite{Stojanovic:21},
is more flexible than the transmon-based one in that it allows one to independently tune the dimensionless
P and BM coupling strengths [cf. Eq.~\eqref{ExplicitFormH_PBM}]. This realization entails two ground states 
of Rb$^{87}$ (i.e., two hyperfine sublevels of its electronic ground state) and two high-lying Rydberg 
states; each of the ground states
is coupled with its corresponding Rydberg state via an off-resonant (dressing) laser (note that without 
a significant loss of generality the system can also be realized with a single dressing laser, in which 
case the two relevant detunings are the same). Thus, instead of interacting through the conventional resonant 
dipole-dipole interaction that amounts to an exchange of two Rydberg states between the two atoms (as 
would be the case in the absence of Rydberg dressing), the resulting Rydberg-dressed interaction is
equivalent to an exchange between two Rydberg-dressed states (i.e. two different linear combinations 
of a ground state and its corresponding Rydberg state, with a small admixture of the latter). When 
the nearly-harmonic optical dipole-trap potential is quantized into Einstein-type bosons, this system 
is effectively described by an itinerant Rydberg-dressed excitation interacting with these bosons via 
P and BM coupling mechanisms. The principal experimental knob in this system is the Rabi frequency of 
the dressing laser. 

In the two described realizations of the model under considerations with interacting qubit arrays, the 
bare-excitation Bloch states -- obtained for $g_{\textrm{P}}=g_{\textrm{BM}}$ (as discussed in Sec.~\ref{BareExcEigenstate} 
above) -- correspond to $N$-qubit $W$ states. Generally speaking, a bare-excitation state $|\Psi_{k}\rangle
\equiv c^{\dagger}_{k}|0\rangle_{\textrm{e}}$ with quasimomentum $k$ corresponds -- via the Jordan-Wigner 
transformation from spinless-fermion to (pseudo)spin-$1/2$ (qubit) degrees of freedom~\cite{ColemanBOOK} -- to 
the generalized (twisted) $W$ state~\cite{Haase++:22}
\begin{equation}\label{defTwistW}
|W_N(k)\rangle = \frac{1}{\sqrt{N}}\sum_{n=1}^{N}
e^{ikn}|0\ldots \underbrace{1}_n \ldots 0\rangle \:,
\end{equation}
a maximally-entangled $N$-qubit state given by an equal superposition of states in which exactly one 
qubit is in the state $|1\rangle$ (with all the remaining qubits being in the state $|0\rangle$). In particular,
the state $|\Psi_{k=0}\rangle\equiv c^{\dagger}_{k=0}|0\rangle_{\textrm{e}}$, which can be realized in 
both physical platform discussed above, corresponds to the conventional $N$-qubit $W$ state~\cite{StojanovicPRL:20,Zhang++:23},
which is completely symmetric with respect to permutations of qubits~\cite{Stojanovic+Nauth:22,Stojanovic+Nauth:23}. 
At the same time, the state $|\Psi_{k=\pi}\rangle\equiv c^{\dagger}_{k=\pi}|0\rangle_{\textrm{e}}$, which can be 
realized in the neutral-atom platform, corresponds to the $\pi$-twisted $W$ state~\cite{Stojanovic:21}.

\section{Entanglement spectrum: Basic aspects} \label{entspectrum}
As a preparation for further considerations, a brief general introduction into entanglement spectra 
and entropy of bipartite quantum systems is presented below (Sec.~\ref{EntSpectrum}). This is followed 
by essential details of the application of these concepts to the coupled e-ph system at hand 
(Sec.~\ref{EntSpectEPH}). 

\subsection{Entanglement spectrum: General considerations}\label{EntSpectrum}
Consider a quantum system that consists of two subsystems $A$ and $B$; its Hilbert space is given 
by the tensor product $\mathcal{H}=\mathcal{H}_{\textrm{A}}\otimes\mathcal{H}_{\textrm{B}}$, where 
$\mathcal{H}_{\textrm{A}}$ ($\mathcal{H}_{\textrm{B}}$) is the Hilbert space of the subsystem $A$ ($B$). 
Let $\{|\mathcal{S}_{\textrm{A}}\rangle\:|\:\mathcal{S}_{\textrm{A}}=1,\ldots,D_{\textrm{A}}\}$ and 
$\{|\mathcal{S}_{\textrm{B}}\rangle\:|\:\mathcal{S}_{\textrm{B}}=1,\ldots,D_{\textrm{B}}\}$ be orthonormal 
bases of $\mathcal{H}_{\textrm{A}}$ and $\mathcal{H}_{\textrm{B}}$, where $D_{\textrm{A}}$ and 
$D_{\textrm{B}}$ are the respective dimensions of these two Hilbert spaces.

An arbitrary pure quantum state in the Hilbert space $\mathcal{H}=\mathcal{H}_{\textrm{A}}\otimes
\mathcal{H}_{\textrm{B}}$ of the bipartite system can be decomposed in the orthonormal basis 
$\{|\mathcal{S}_{\textrm{A}}\rangle \otimes |\mathcal{S}_{\textrm{B}}\rangle\}$, i.e.
\begin{equation}\label{statePsi}
|\Psi\rangle = \sum_{\mathcal{S}_{\textrm{A}}=1}^{D_{\textrm{A}}}\sum_{\mathcal{S}_{\textrm{B}}=1}
^{D_{\textrm{B}}}\:C_{\mathcal{S}_{\textrm{A}},\mathcal{S}_{\textrm{B}}}\:
|\mathcal{S}_{\textrm{A}}\rangle\otimes|\mathcal{S}_{\textrm{B}}\rangle \:,
\end{equation}
where the coefficients $C_{\mathcal{S}_{\textrm{A}},\mathcal{S}_{\textrm{B}}}$ in this expansion can be 
viewed as the entries of a certain matrix. By means of the singular-value decomposition (SVD), this 
matrix -- which will hereafter be denoted by $M$ and referred to as the {\em entanglement matrix} -- can 
be written as 
\begin{equation}\label{SVDentMatrix}
M = U\mathcal{D}V^{\dagger} \:.
\end{equation}
Here $U$ is a matrix of dimension $D_{\textrm{A}}\times \textrm{min} (D_{\textrm{A}},D_{\textrm{B}})$ that satisfies 
the condition $U^{\dagger}U = \mathbbm{1}$ and $V$ a $D_{\textrm{B}}\times \textrm{min}(D_{\textrm{A}},D_{\textrm{B}})$ 
matrix with the property that $VV^{\dagger}=\mathbbm{1}$; $\mathcal{D}$ is a diagonal square matrix of dimension 
$\textrm{min}(D_{\textrm{A}},D_{\textrm{B}})$, whose matrix elements -- the singular values of $M$ -- are non-negative. 
Therefore, these singular values can be written in the form $e^{-\xi_{\alpha}/2}$, where $\alpha=1,\ldots,\textrm{min} 
(D_{\textrm{A}},D_{\textrm{B}})$.

The density matrix corresponding to a pure state $|\Psi\rangle$ in the Hilbert space 
$\mathcal{H}=\mathcal{H}_{\textrm{A}}\otimes\mathcal{H}_{\textrm{B}}$ is given by
\begin{equation}\label{rhogen}
\rho=\frac{|\Psi\rangle\langle\Psi|}{\langle\Psi|\Psi\rangle}  \:.
\end{equation}
The reduced (marginal) density matrix $\rho_{\textrm{\tiny{A}}}$ of the subsystem 
$A$ is obtained by tracing $\rho$ over the degrees of freedom of the subsystem $B$: 
$\rho_{\textrm{A}}=\tr_{\textrm{B}}\rho$. [Analogously, the reduced density matrix 
$\rho_{\textrm{\tiny{B}}}$ of the subsystem $B$ is given by $\rho_{\textrm{B}}=
\tr_{\textrm{A}}\rho$.] Using the SVD [cf. Eq.~\eqref{SVDentMatrix}], one obtains 
the Schmidt decomposition of the generic pure state $|\Psi\rangle$~\cite{Schmidt:1907,Ekert+Knight:95}
\begin{equation}\label{SchmidtDecomp}
|\Psi\rangle = \sum_{\alpha=1}^{\alpha_{\textrm{max}}}\:e^{-\xi_{\alpha}/2}
|\psi^{\alpha}_\textrm{A}\rangle\otimes|\psi^{\alpha}_\textrm{B}\rangle \:,
\end{equation}
where $\alpha_{\textrm{max}}\equiv\textrm{min}(D_{\textrm{A}},D_{\textrm{B}})$, and
\begin{equation} \label{psiAB}
|\psi^{\alpha}_{\mu}\rangle = \sum_{s_{\mu}=1}^{d_{\mu}}\:
U^{\dagger}_{\alpha,s_{\mu}}|\mathcal{S}_{\mu}\rangle \:\quad (\:\mu = A,B\:)\:,
\end{equation}
are the singular vectors of the matrix $M$. By making use of the Schmidt decomposition 
of the state $|\Psi\rangle$ (or, equivalently, the SVD of the entanglement matrix), 
the two reduced density matrices can jointly be written in the spectral form
\begin{eqnarray} \label{rhoArhoB}
\rho_{\mu} = \sum_{\alpha=1}^{\alpha_{\textrm{max}}}\:e^{-\xi_{\alpha}}|\psi^{\alpha}_{\mu} 
\rangle\langle\psi^{\alpha}_{\mu}| \:\quad (\:\mu = A,B\:)\:.
\end{eqnarray}
This form renders it manifest that the joint eigenvalues of these two reduced density 
matrices are given by the squares $e^{-\xi_{\alpha}}$ of the above singular values
of the entanglement matrix. 

The very notion of the entanglement spectrum originates from the fact that each reduced 
density matrix can be writen in the form $\exp(-H_{\textrm{E}})$, this being 
the canonical density matrix that pertains to the Hamiltonian $H_{\textrm{E}}$ at the 
inverse temperature $\beta_{\textrm{E}}=1$~\cite{Chandran+:14}. The Hamiltonian 
$H_{\textrm{E}}$ -- the negative logarithm of the reduced density matrix of the system -- is 
usually referred to as the entanglement (or modular) Hamiltonian and the entanglement spectrum 
is given by its set of eigenvalues. In particular, the entanglement spectrum of the bipartite system 
whose reduced density matrices are given by Eq.~\eqref{rhoArhoB} is the set $\{\xi_{\alpha}\:|
\:\alpha =1,\ldots,\textrm{min}(D_{\textrm{A}},D_{\textrm{B}})\}$ of the negative logarithms 
of the joint eigenvalues $e^{-\xi_{\alpha}}$ of $\rho_{\textrm{A}}$ and $\rho_{\textrm{B}}$. 

Having defined the entanglement spectrum, it is of interest to establish its connection 
with the entanglement entropy, which can -- generally speaking -- be thought of as the 
thermodynamic entropy of a system governed by the entanglement Hamiltonian $H_{\textrm{E}}$~\cite{Wehrl:78,HayashiBOOK}. 
For the above generic bipartite system this entropy is given by
\begin{equation} \label{vonNeumannS}
S_{E}= -\textrm{Tr}_{\textrm{A}}(\rho_{\textrm{A}}
\ln\rho_{\textrm{A}}) 
\end{equation}
and can equivalently be expressed as $S_{E}=-\tr_{\textrm{B}}(\rho_{\textrm{\tiny{B}}}
\ln\rho_{\textrm{B}})$. By making use of the spectral form of the operator $\rho_{\textrm{A}}$
[cf. Eq.~\eqref{rhoArhoB}] and the fact that $\rho_{\textrm{A}}\ln\rho_{\textrm{A}}$ is an 
analytic operator function of $\rho_{\textrm{A}}$, one can straghtforwardly express $S_{E}$
in terms of the entanglement-spectrum eigenvalues $\xi_{\alpha}$:
\begin{equation} \label{EntangleSvsEntSpect}
S_{E}=\sum_{\alpha=1}^{\textrm{min}(D_{\textrm{A}},D_{\textrm{B}})}
\xi_{\alpha}e^{-\xi_{\alpha}} \:.
\end{equation}

%The interpretation of entanglement spectra is facilitated by the following property of
%the reduced density matrices. Let $\mathcal{O}$ be a Hermitian operator (an observable)
%that acts on the tensor-product Hilbert space $\mathcal{H}=\mathcal{H}_{\textrm{A}}
%\otimes\mathcal{H}_{\textrm{B}}$. Assume that this operator can be written as a sum $\mathcal{O}=\mathcal{O}_{\textrm{A}}
%+\mathcal{O}_{\textrm{B}}$ of two operators, with $\mathcal{O}_{\textrm{A}}$ acting exclusively
%on the Hilbert space $\mathcal{H}_{\textrm{A}}$ of the subsystem $A$ and $\mathcal{O}_{\textrm{B}}$ 
%only on the Hilbert space $\mathcal{H}_{\textrm{B}}$ of the subsystem $B$. Assuming that the state 
%$|\Psi\rangle$ is an eigenstate of $\mathcal{O}$ -- which immediately implies that its corresponding 
%density matrix $\rho$ commutes with $\mathcal{O}$ -- it can straightforwardly be demonstrated that 
%$\mathcal{O}_{\textrm{A}}$ commutes with the reduced density matrix $\rho_{\textrm{A}}$~\cite{Stojanovic:20}. 
%Consequently, the operators $\rho_{\textrm{A}}$ and $\mathcal{O}_{\textrm{A}}$ can simultaneously 
%be diagonalized -- i.e., these two operators have a joint eigenbasis -- and the entanglement-spectrum 
%eigenvalues can be labelled using the quantum numbers associated with $\mathcal{O}_{\textrm{A}}$.

\subsection{Entanglement spectrum and entropy of the coupled e-ph system}\label{EntSpectEPH}
The Hilbert space of the coupled e-ph system under consideration is given by the tensor-product 
Hilbert space ${\mathcal H}={\mathcal H}_{\textrm{e}}\otimes{\mathcal H}_{\textrm{ph}}$ of the 
excitation- (${\mathcal H}_{\textrm{e}}$) and phonon (${\mathcal H}_{\textrm{ph}}$) spaces,
their respective dimensions being denoted by $D_{\textrm{e}}$ and $D_{\textrm{ph}}$ in the following. 
By making use of Eq.~\eqref{rhogen}, the density matrix corresponding to the ground state 
$|\psi_{\textrm{gs}}\rangle$ of system can be expressed as
\begin{equation}\label{rho_eph}
\rho^{(\textrm{gs})}_{\textrm{e-ph}}=\frac{|\psi_{\textrm{gs}}\rangle
\langle\psi_{\textrm{gs}}|}{\langle\psi_{\textrm{gs}}|\psi_{\textrm{gs}}\rangle} \:.
\end{equation}
The reduced excitation density matrix is then obtained by tracing out the phonon 
degrees of freedom:
\begin{equation} \label{rho_e}
\rho^{(\textrm{gs})}_{\textrm{e}}= \textrm{Tr}_{\textrm{ph}}
\big[\rho^{(\textrm{gs})}_{\textrm{e-ph}}\big] \:.
\end{equation}
The explicit derivation of this reduced density matrix (i.e. of its entries), based 
on the use of a symmetry-adapted basis of the Hilbert space of the system (see 
Sec.~\ref{CompStrategy} below), is presented in Appendix~\ref{RedDensMatEl}.

As a special case of Eq.~\eqref{vonNeumannS}, the ground-state entanglement entropy
$S^{(\textrm{gs})}_{E}$ of the system is given by
\begin{equation} \label{S_gs}
S^{(\textrm{gs})}_{E}= -\textrm{Tr}_{\textrm{e}}\big[\rho^{(\textrm{gs})}_{\textrm{e}}
\ln\rho^{(\textrm{gs})}_{\textrm{e}}\big] \:.
\end{equation}
For a coupled e-ph system defined on a discrete lattice with $N$ sites, the dimension of 
the excitation Hilbert space is $N$, while the dimension of the phonon Hilbert space is 
much larger than that (see Sec.~\ref{CompStrategy} below). Thus, $\alpha_{\textrm{max}}
\equiv\textrm{min}(D_{\textrm{e}},D_{\textrm{ph}})=N$ [cf. Sec.~\ref{EntSpectrum}] and 
the ground-state entanglement spectrum of the system at hand consists of $N$ eigenvalues.
Based on the general expression in Eq.~\eqref{EntangleSvsEntSpect}, the ground-state 
entanglement entropy of this system [cf. Eq.~\eqref{S_gs}] can be expressed as 
\begin{equation} \label{EntangleEntropyEPH}
S^{(\textrm{gs})}_{E}=\sum_{\alpha=1}^{N}\xi^{(\textrm{gs})}_{\alpha}
e^{-\xi^{(\textrm{gs})}_{\alpha}} \:.
\end{equation}

\section{Hilbert-space truncation and symmetry-adapted basis} \label{CompStrategy}
In this section, the structure of the Hilbert space of the system and its controlled
truncation are first discussed (Sec.~\ref{HilbSpaceTrunc}). This is followed by the 
introduction of the symmetry-adapted basis of this Hilbert space (Sec.~\ref{SAB}).

\subsection{Hilbert space and its truncation}   \label{HilbSpaceTrunc}
Because phonon Hilbert spaces are infinite-dimensional, the treatment of the coupled e-ph 
system at hand requires a controlled truncation of the phonon Hilbert space. 
The Hilbert space of the system at hand, defined on a 1D lattice with $N$ sites, is spanned by states 
$|n\rangle_e\otimes|\mathbf{m}\rangle_\text{ph}$, where $|n\rangle_e\equiv c_{n}^{\dagger}|0\rangle_e$ 
is the state with the excitation located at site $n$ ($n=1,\ldots,N$) and $|\mathbf{m}\rangle_\text{ph}$
is a phonon Fock state. This state is given by
\begin{equation}\label{mphvect}
|\mathbf{m}\rangle_\text{ph} = \prod_{n=1}^{N\otimes}
\frac{(a_n^\dagger)^{m_n}}{\sqrt{m_n!}}\:|0\rangle_\text{ph}\:,
\end{equation}
where $\mathbf{m}\equiv (m_1,\ldots,m_N)$ and $m_n$ is the phonon occupation number at site $n$. 

Given the infinite-dimensional nature of phonon Hilbert spaces, one has to restrict oneself 
to the truncated phonon Hilbert space that consists of states with the total phonon number 
$m=\sum_{n=1}^N m_n$ (where $0\le m_n \le m$) not larger than a certain maximal value $N_\text{ph}$. 
Accordingly, the dimension of the truncated phonon Hilbert space is $D_\text{ph}=(N_\text{ph}+N)!/
(N_\text{ph}!N!)$. 

It should be stressed that the chosen values for the number of sites $N$ (i.e., the system size)
and the maximal total number $N_\text{ph}$ of phonons, which jointly determine the total dimension 
of the truncated Hilbert space of the system, is dictated by the required accuracy of computing the 
sought-after physical observables [for instance, the ground-state energy of the system, the expected 
phonon number in the ground state, the quasiparticle residue (spectral weight), etc.]. The actual 
truncation of the Hilbert space of the coupled e-ph system is performed through a gradual increase 
of the number of sites $N$ -- combined with an increase in the total number of phonons 
$N_\text{ph}$ -- up to the point where a further increase of $N$ and $N_\text{ph}$ does not cause 
an appreciable change (with respect to a pre-defined error margin) in the obtained numerical results 
for the desired physical quantities. 

\subsection{Symmetry-adapted basis}    \label{SAB}
Because the dimension of the excitation Hilbert space is equal to $N$, 
the dimension of the total Hilbert space of the coupled e-ph system under consideration is given 
by $D_\text{e-ph} = N \times (N_\text{ph}+N)!/(N_\text{ph}!N!)$. However, the actual dimension of 
the matrix-diagonalization problem for the total system Hamiltonian can be additionally reduced 
by exploiting the discrete translational symmetry of the system, which is mathematically expressed 
as the commutation $[H,K_{\mathrm{tot}}]=0$ of the operators $H$ and $K_{\mathrm{tot}}$. 

The explicit use of the discrete traslational symmetry of the system permits the diagonalization 
of $H$ in sectors of the total Hilbert space that correspond to the eigensubspaces of $K_{\mathrm{tot}}$; 
each of those sectors has the dimension equal to that of the truncated phonon space (i.e., $D_{K}=
D_{\textrm{ph}}$). Therefore, it is natural to make use of the symmetry-adapted basis, which for 
fixed $K\in (-\pi,\pi]$ and different phonon Fock states $|\mathbf{m}\rangle_\text{ph}$ is given by  
\begin{equation}\label{symmbasis}
|K,\mathbf{m}\rangle = N^{-1/2} \sum_{n=1}^N e^{iKn}\,\mathcal{T}_{n-1}(|1\rangle_\text{e} 
\otimes |\mathbf{m}\rangle_\text{ph}) \:,
\end{equation}
where $\mathcal{T}_{n}$ ($n=0,1,\ldots, N-1$) are the (discrete) translation operators. 
The last equation can straightforwardly be recast as
\begin{equation}\label{symmbasalter}
|K,\mathbf{m}\rangle = N^{-1/2} \sum_{n=1}^N e^{iKn}\, |n\rangle_\text{e} 
\otimes \mathcal{T}^{\textrm{ph}}_{n-1}|\mathbf{m}\rangle_\text{ph} \:,
\end{equation}
where the discrete-translation operators $\mathcal{T}^{\textrm{ph}}_{n-1}$ act on the phonon 
Hilbert space. In particular, the $l$-th occupation number corresponding to the state 
$|\mathcal{T}^{\textrm{ph}}_{n-1}\mathbf{m}\rangle$ is given by $m_{s(l,n-1)}$, where
\begin{equation}\label{findef}
s(l,n)\equiv\begin{cases} 
N-n+l\:, \:\text{for $l\le n$} \\
l-n\:, \: \text{for $l>n$}
\end{cases}\:.
\end{equation}
An arbitrary state in the fixed-$K$ sector of the Hilbert space of the 
coupled e-ph system at hand can be expressed as a linear combination of
the states in Eq.~\eqref{symmbasalter}. In particular, the eigenstates 
$|\psi^{(i)}_{K}\rangle$ ($i=1,\ldots,D_\text{ph}$) of total Hamiltonian $H$ 
that correspond to the value $K$ of the total quasimomentum operator 
can be written as 
\begin{equation}\label{decomp_genpsi}
|\psi^{(i)}_{K}\rangle=\sum_{\mathbf{m}}C^{(i)}_{K,\mathbf{m}}
|K,\mathbf{m}\rangle \:.
\end{equation}
As a special case of Eq.~\eqref{decomp_genpsi}, the ground state $|\psi_{\textrm{gs}}\rangle$ 
of the system, which belongs to the $K=K_{\textrm{gs}}$ Hilbert-space sector, can be expanded 
in the symmetry-adapted basis as 
\begin{equation}\label{decomp_psigs}
|\psi_{\textrm{gs}}\rangle=\sum_{\mathbf{m}}C_{K_{\textrm{gs}},\mathbf{m}}
|K_{\textrm{gs}},\mathbf{m}\rangle \:.
\end{equation}

The use of the symmetry-adapted basis allows one to significantly alleviate
the computational burden involved in the exact-diagonalization treatment of 
the coupled e-ph system under consideration. Instead of carying out an exact 
diagonalization of a $D_\text{e-ph}\times D_\text{e-ph}$ matrix, it suffices
to perform $N$ diagonalizations of $(D_\text{e-ph}/N) \times (D_\text{e-ph}/N)$
matrices.

\section{Results and Discussion} \label{ResultsDiscuss}
Following a short summary of the parameters values used in the numerically-exact 
evaluation of the entanglement spectrum and entropy of the coupled e-ph system under 
consideration (Sec.~\ref{EDdetails}), the principal findings of the present work are 
presented and discussed below (Sec.~\ref{GenericCase}).

\subsection{Evaluation of the entanglement spectrum}      \label{EDdetails}
The numerically-exact evaluation of the ground-state entanglement spectrum of the coupled e-ph model 
under consideraton consists of the following three steps. 

Firstly, the ground-state vector $|\psi_{\textrm{gs}}\rangle$ -- represented by the expansion coefficients 
$C_{K_{\textrm{gs}},\mathbf{m}}$ in the symmetry-adapted basis [cf. Eq.~\eqref{decomp_psigs}] -- is 
obtained by means of Lanczos diagonalization~\cite{CullumWilloughbyBook,PrelovsekBoncaChapter:13} of the total Hamiltonian 
$H=H_0+H_{\textrm{P}}+H_{\textrm{BM}}$ of the system [cf. Eqs.~\eqref{H_0} and \eqref{ExplicitFormH_PBM}].
The ground state is determined after a controllable truncation of the Hilbert space of the system based on 
the scheme described in Sec.~\ref{HilbSpaceTrunc}; the adopted convergence criterion was that the relative error 
in the ground-state energy and the phonon distribution is not larger than $10^{-4}$. For the system at hand,
it was verified that this criterion is satisfied for a system with $N=8$ sites (with the periodic boundary conditions) 
and the maximal total number of $N_\text{ph}=9$ phonons. Therefore, the ground state was evaluated for 
an eight-site ring and with the phonon Hilbert space of dimension $D_\text{ph}=24,310$.

Secondly, having computed $|\psi_{\textrm{gs}}\rangle$, the reduced density matrix $\rho^{(\textrm{gs})}_{\textrm{e}}$ is obtained
using Eqs.~\eqref{rho_eph} and \eqref{rho_e}. Its matrix elements $(\rho^{(\textrm{gs})}_{\textrm{e}})_{nn'}$ ($n,n'=1,\ldots,N$) 
are given by (for a detailed derivation, see Appendix~\ref{RedDensMatEl})
\begin{eqnarray}\label{rhoematrel}
\left(\rho^{(\textrm{gs})}_{\textrm{e}}\right)_{nn'}&=& N^{-1}\:e^{iK_{\textrm{gs}}(n-n')}\sum_{\mathbf{m},\mathbf{m'}}
\:C^{*}_{K_{\textrm{gs}},\mathbf{m'}}\:C_{K_{\textrm{gs}},\mathbf{m}}\nonumber \\
&\times&\:\langle\mathbf{m'}\:|\:\mathcal{T}^{\textrm{ph}}_{n-n'}\mathbf{m}\rangle\:,
\end{eqnarray}
The matrix element $\langle\mathbf{m'}\:|\:\mathcal{T}^{\textrm{ph}}_{n-n'}\mathbf{m}\rangle$ in the last 
equation is computed by making use of Eq.~\eqref{findef}, noting at the same time that $\langle\mathcal{T}^{\textrm{ph}}
_{n'}\mathbf{m'}\:|\:\mathcal{T}^{\textrm{ph}}_{n}\mathbf{m}\rangle=1$ if all the corresponding phonon 
occupation numbers in $|\mathcal{T}^{\textrm{ph}}_{n}\mathbf{m}\rangle$ and $|\mathcal{T}^{\textrm{ph}}_{n'}
\mathbf{m'}\rangle$ are equal; otherwise, this matrix element is equal to zero.

Finally, once the matrix elements of $\rho^{(\textrm{gs})}_{\textrm{e}}$ have been obtained, the $N$ entanglement-spectrum eigenvalues 
and their corresponding eigenvectors -- for each fixed value of the effective e-ph coupling strength -- are 
determined by solving the ($N\times N$)-dimensional eigenvalue problem of $\rho^{(\textrm{gs})}_{\textrm{e}}$. The ground-state 
entanglement entropy is then straightforwardly obtained from the computed entanglement-spectrum eigenvalues using 
Eq.~\eqref{EntangleSvsEntSpect}. 

\subsection{Results for the ground-state entanglement spectrum and entropy}   \label{GenericCase}
It is pertinent to discuss the entanglement-related properties of the model at hand by exploring the entire range 
of e-ph coupling strengths. In other words, both the weak-coupling regime -- characterized by a quasi-free (weakly phonon-dressed)
excitation -- and its strong-coupling counterpart where a heavily-dressed excitation (small polaron) is formed, 
along with the intermediate regime, will be discussed in what follows. The analysis presented below will also 
include different values of the adiabaticity ratio -- both the adiabatic ($\omega_{\textrm{ph}}/t_{\rm e}<1$) and 
antiadiabatic ($\omega_{\textrm{ph}}/t_{\rm e}>1$) regimes, as well as the case with $\omega_{\textrm{ph}}/t_{\rm e}=1$. 

Given that -- by contrast to the BM coupling -- the P coupling itself allows the occurrence of sharp ground-state 
transitions (i.e. nonanalyticities in the relevant quantities) it is pertinent to perform the analysis of the model 
under consideration by varying the P-coupling strength in the presence of BM coupling of fixed strength. In what follows,
the effective P-coupling strength [cf. Eq.~\eqref{lambdaPandBM}] will be varied from $\lambda_{\textrm{P}}=0$ to 
$\lambda_{\textrm{P}}=3.2$, with two fixed BM-coupling strengths ($g_{\textrm{BM}}=0.25$ and $g_{\textrm{BM}}=0.4$).

\begin{figure}[t!]
\includegraphics[clip,width=8.5cm]{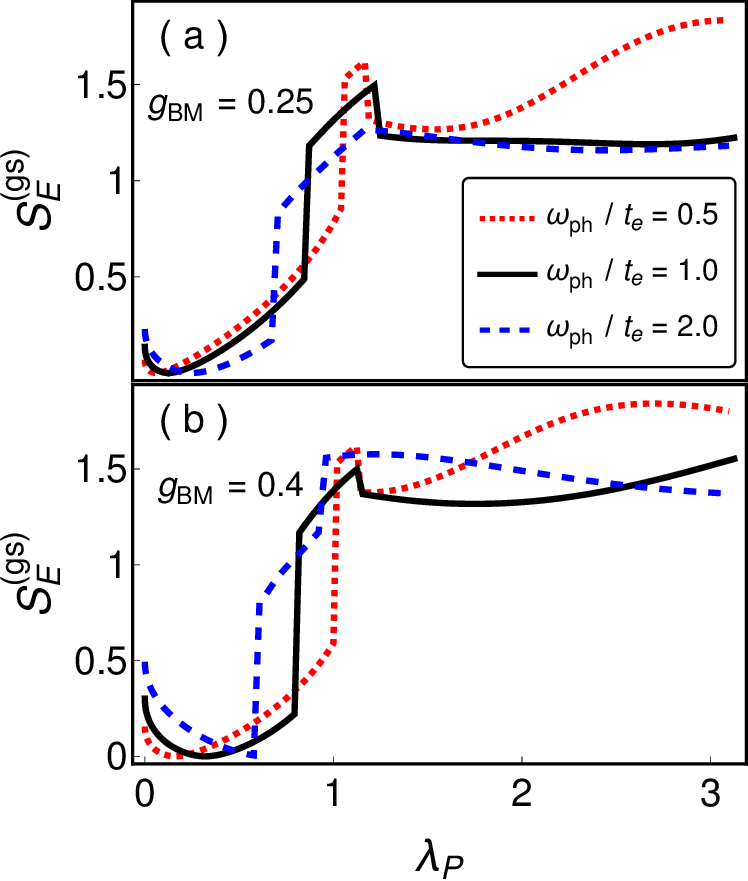}
\caption{\label{fig:EntropyPlot}Ground-state e-ph entanglement entropy as a function of 
the effective P-coupling strength, depicted for three different values of the adiabaticity 
ratio and two fixed values of the BM coupling strength $g_{\textrm{BM}}$: (a) $g_{\textrm{BM}}=0.25$, 
and (b) $g_{\textrm{BM}}=0.4$.}
\end{figure}

Prior to discussing in detail the ground-state entanglement spectrum of the model at hand it is instructive to analyze 
the results obtained for its corresponding entanglement entropy $S^{\textrm{(gs)}}_{E}$ [cf. Eq.~\eqref{S_gs}] as a 
function of $\lambda_{\textrm{P}}$. Figure~\ref{fig:EntropyPlot} shows this quantity for three different 
values of the adiabaticity ratio $\omega_{\textrm{ph}}/t_{\rm e}$. The obtained numerical results for $S^{\textrm{(gs)}}_{E}$
show three salient features.

Firstly, the most salient feature of the obtained dependence of $S^{\textrm{(gs)}}_{E}$ on $\lambda_{\textrm{P}}$
is the occurrence of sharp transitions (i.e. first-order nonanalyticities) at certain critical values 
$\lambda^{\textrm{c}}_{\textrm{P}}$ of $\lambda_{\textrm{P}}$. This is a manifestation of the sharp, 
level-crossing transitions characterizing all ground-state-related quantities for models of this type 
[recall the discussion in Sec.~\ref{ModelHamiltonian}]. What can be inferred from Fig.~\ref{fig:EntropyPlot} 
is that for the model at hand there are as many as four such sharp transitions. Another, closel related, 
observation is is that the critical value of $\lambda_{\textrm{P}}$ that corresponds to the first of those
transitions decreases with increasing adiabaticity ratio; in other words, for higher adiabaticity ratios
this transitions occurs for smaller coupling strengths. 

Secondly, the behavior of the entanglement entropy as a function of the effective e-ph coupling strength
is here markedly different from the previously investigated behavior of this quantity in the presence of 
Holstein-type (local)~\cite{Stojanovic:08} or P-type coupling~\cite{Stojanovic:20}. Namely, in those
cases the entanglement entropy grows monotonously with increasing coupling strength, regardless of whether
the strong-coupling regime is characterized by the presence of sharp transitions (as in the case of 
P-type coupling) or just a smooth crossover (the case of Holstein-type coupling); in both cases, this 
entropy saturates at the value $\ln N$ (for the system size $N=8$ discussed here this maximal value is 
around $2.08$), which signifies maximally-entangled states~\cite{Zhao+:04}. Here, by contrast, the entanglement 
entropy shows a nonmonotonic dependence on $\lambda_{\textrm{P}}$ and does not reach the aforementioned
maximal value [cf. Fig.~\ref{fig:EntropyPlot}].

Finally, in the special case of equal P- and BM coupling strengths the entanglement entropy vanishes, 
consistent with the fact that the ground state of the system in this case is a bare-excitation state 
(recall the discussion in Sec.~\ref{BareExcEigenstate}). It is interesting to note that in the case 
with $g_{\textrm{BM}}=0.4$ and the highest value $\omega_{\textrm{ph}}/t_{\rm e}=2$ of the adiabaticity
ratio investigated here, the first sharp transition practically coincides with this special case of
the model [for an illustration, see Fig.~\ref{fig:EntropyPlot}(b)].

\begin{figure}[t!]
\includegraphics[clip,width=8.5cm]{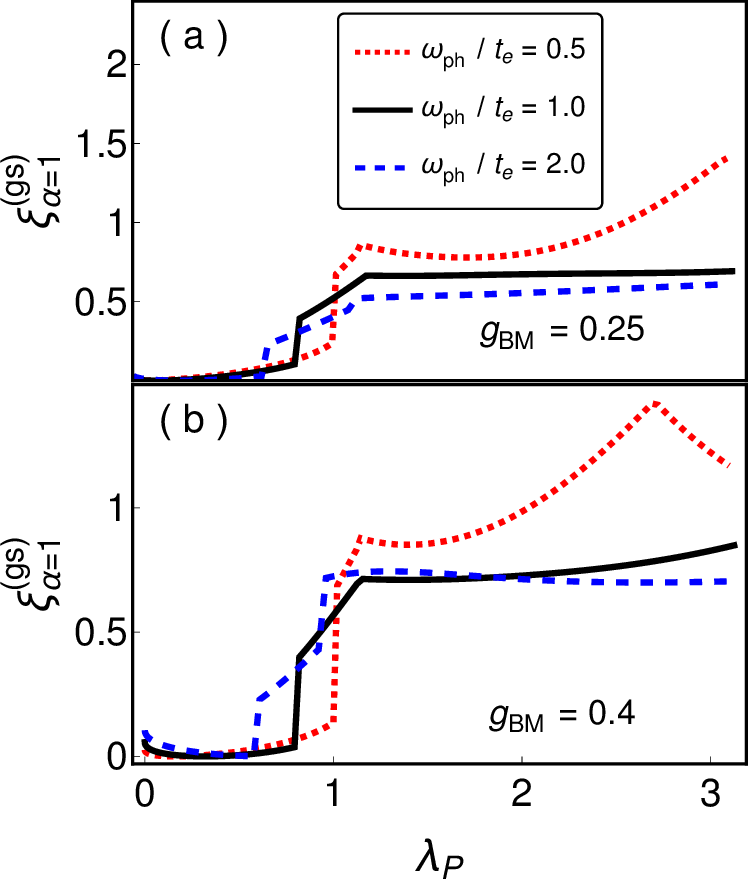}
\caption{\label{fig:xieq1}Dependence of the ground-state entanglement-spectrum eigenvalue 
$\xi^{\textrm{(gs)}}_{\alpha=1}$ on the effective P-coupling strength, shown for three different 
values of the adiabaticity ratio and two fixed BM coupling strengths: 
(a) $g_{\textrm{BM}}=0.25$, and (b) $g_{\textrm{BM}}=0.4$.}
\end{figure}

\begin{figure}[b!]
\includegraphics[clip,width=8.5cm]{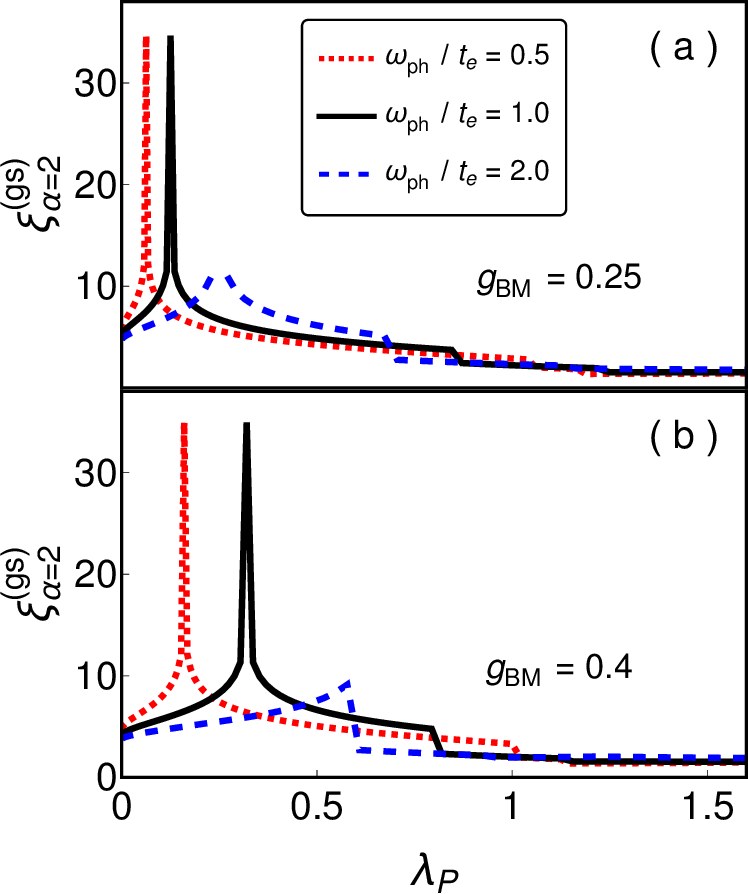}
\caption{\label{fig:xieq2}Dependence of the ground-state entanglement-spectrum eigenvalue 
$\xi^{\textrm{(gs)}}_{\alpha=2}$ on the effective P-coupling strength, shown for three different 
values of the adiabaticity ratio and two fixed BM coupling strengths: 
(a) $g_{\textrm{BM}}=0.25$, and (b) $g_{\textrm{BM}}=0.4$.}
\end{figure}

\begin{figure}[t!]
\includegraphics[clip,width=8.5cm]{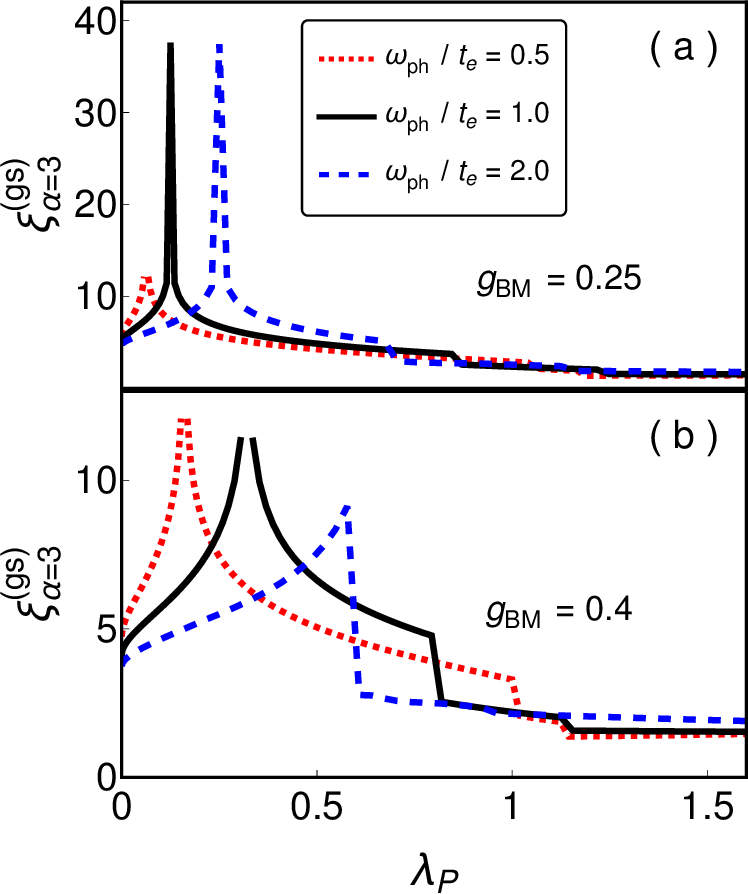}
\caption{\label{fig:xieq3}Dependence of the ground-state entanglement-spectrum eigenvalue 
$\xi^{\textrm{(gs)}}_{\alpha=3}$ on the effective P-coupling strength, shown for three different 
values of the adiabaticity ratio and two fixed BM coupling strengths: 
(a) $g_{\textrm{BM}}=0.25$, and (b) $g_{\textrm{BM}}=0.4$.}
\end{figure}

\begin{figure}[t!]
\includegraphics[clip,width=8.5cm]{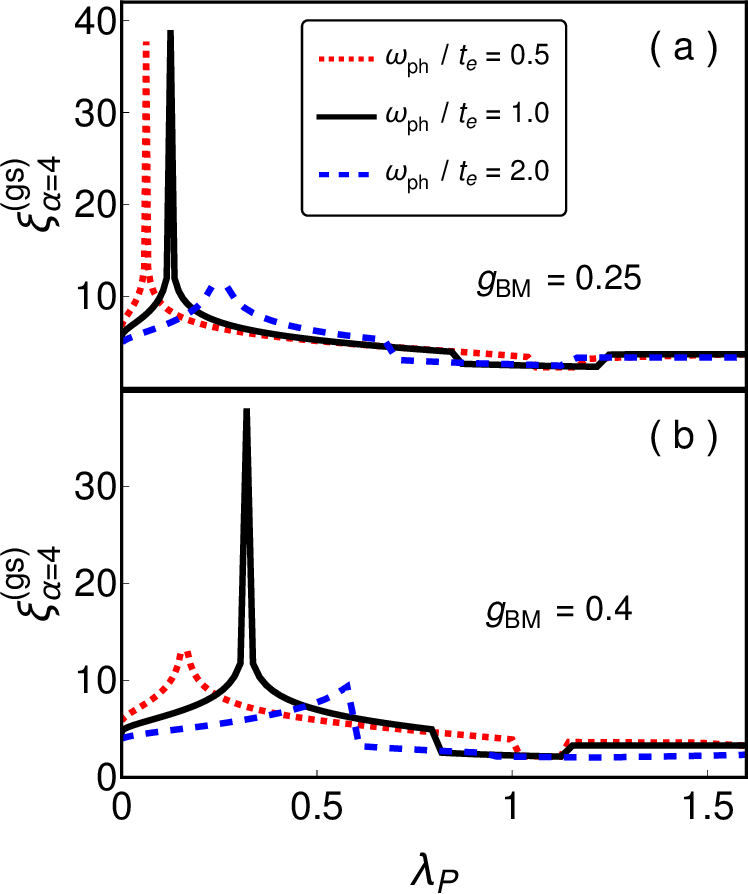}
\caption{\label{fig:xieq4}Dependence of the ground-state entanglement-spectrum eigenvalue 
$\xi^{\textrm{(gs)}}_{\alpha=4}$ on the effective P-coupling strength, shown for three different 
values of the adiabaticity ratio and two fixed BM coupling strengths: 
(a) $g_{\textrm{BM}}=0.25$, and (b) $g_{\textrm{BM}}=0.4$.}
\end{figure}

Having considered the gross features of the e-ph entanglement in the model at hand -- as described 
by the entanglement entropy -- the more subtle features can be analyzed through the prism of the 
corresponding entanglement spectrum. The entanglement-spectrum eigenvalues 
$\xi^{\textrm{(gs)}}_{\alpha}$ ($\alpha = 1,2,\ldots,8$) -- i.e., their dependence on 
$\lambda_{\textrm{P}}$ -- are depicted in Figs.~\ref{fig:xieq1} -- \ref{fig:xieq4} for 
$\alpha=1,2,3,4$, respectively. The most apparent feature of these eigenvalues is that they
reflect the presence of multiple sharp transitions in the ground state of the model under 
consideration. Another relevant observation is that the qualitative structure of this spectrum 
does not display a strong dependence on the adiabaticity ratios, i.e. it is fairly similar for 
different values of $\omega_{\textrm{ph}}/t_{\rm e}$. This conclusion bears some resemblance to
the previously established general properties of phonon-dressed excitations formed in the presence 
of P-type e-ph coupling~\cite{Capone+:97}. 

What can be inferred from Fig.~\ref{fig:xieq1}, which shows the smallest eigenvalue 
$\xi^{\textrm{(gs)}}_{\alpha=1}$ in the ground-state entanglement spectrum, is that the 
behavior of $\xi^{\textrm{(gs)}}_{\alpha=1}$ largely mimics the beavior of the entanglement
entropy itself. In other words, the dependence of $\xi^{\textrm{(gs)}}_{\alpha=1}$ on 
$\lambda_{\textrm{P}}$ is qualitatively similar to that of $S^{\textrm{(gs)}}_{E}$. In
particular, this smallest eigenvalue also vanishes -- like $S^{\textrm{(gs)}}_{E}$ 
itself -- in the special case $g_{\textrm{P}}=g_{\textrm{BM}}$ of the model. 

The last conclusion -- namely, that the ground-state entanglement entropy $S^{\textrm{(gs)}}_{E}$ is to a large extent 
determined by the smallest entanglement-spectrum eigenvalue $\xi^{\textrm{(gs)}}_{\alpha=1}$ -- is in accordance with 
findings of related studies of other many-body systems. To be more specific, it was already observed that the universal 
part of the entanglement spectrum is typically determined predominantly by the largest eigenvalues of the corresponding 
reduced density matrix~\cite{Johri+:17}. Moreover, given that the entanglement entropy that corresponds to a certain 
reduced density matrix is equivalent to the thermodynamic entropy of the  
attendant entanglement Hamiltonian $H_{\textrm{E}}$ at the inverse temperature $\beta_{\textrm{E}}=1$ [recall the discussion in 
Sec.~\ref{EntSpectrum}], the last finding has another important implication. Namely, this finding is intimately related to 
the quite general issue as to whether the Hamiltonian of a generic many-body system can be thought of as being essentially
encoded in a single eigenstate (e.g., its ground state). Such situations have so far been discussed only 
in the context of thermodynamic and entanglement entropies of single-component systems [for instance, coupled quantum 
spin-$1/2$ chains or interacting hard-core bosons in 1D systems~\cite{Garrison+Grover:18}]. Thus, the present study 
of the entanglement spectrum of a coupled e-ph model provides a qualitatively dissimilar instance of an interacting 
system in which the same issue is of interest.

By contrast to the smallest eigenvalue, all the remaining entanglement-spectrum eigenvalues show qualitatively 
similar behavior as a function of $\lambda_{\textrm{P}}$. The only common feature of their dependence on $\lambda_{\textrm{P}}$
with that of $\xi^{\textrm{(gs)}}_{\alpha=1}$ is the fact that they also display the nonanalytic behavior manifesting 
the aforementioned sharp transitions. On the other hand, their behavior in the special case $g_{\textrm{P}}=g_{\textrm{BM}}$ 
of the model is drastically different than that of $\xi^{\textrm{(gs)}}_{\alpha=1}$. Namely, as can be inferred from 
Figs.~\ref{fig:xieq2} -- \ref{fig:xieq4} (other eigenvalues are not shown so as to avoid redundancy) all those eigenvalues
display a singularity -- i.e., diverge -- in this special case of the model. However, given that $x e^{-x}\rightarrow 0$
as $x \rightarrow \infty$, all those eigenvalues still give vanishing contributions to the ground-state entanglement 
entropy in this special case of the model [cf. Eq.~\eqref{EntangleEntropyEPH}]; in other words, $\xi^{(\textrm{gs})}_{\alpha}
e^{-\xi^{(\textrm{gs})}_{\alpha}}\rightarrow 0$ for $\alpha=2,\ldots,8$ in this special case.

As already mentioned in Sec.~\ref{PhysRealize}, in the special case $g_{\textrm{P}}=g_{\textrm{BM}}$ 
of the model under consideration the ground state below a critical coupling strength corresponds -- after
performing the Jordan-Wigner transformation -- to a $W$ state. What makes the prospect of realizing  
$W$ states in the two proposed analog simulators of the model with simultaneous P and BM coupling
~\cite{Stojanovic:21,StojanovicPRL:20} particularly appealing is the fact that this $W$ state is the 
actual ground state of the system in a parametrically large window of the relevant physical parameters, 
which is quite a rare occurrence in physical platforms for quantum computing. As a result, the envisioned 
realizations of $W$ states can be expected to be extremely robust.

One important aspect of the envisioned realizations of multipartite $W$ states as ground state of 
the model under consideration pertains to the entanglement between an itinerant excitation and bosonic degrees 
of freedom in the system -- microwave photons in the resonators in the superconducting-qubit-based proposal 
of Ref.~\cite{StojanovicPRL:20} and quanta of vibrations in harmonic microtraps in the envisioned 
neutral-atom-based realizations of Ref.~\cite{Stojanovic:21}. The excitation-boson entanglement in 
those systems -- which is quantitatively described by the findings of the present study -- represents
a source of (boson-induced) decoherence if the condition of equality of the relevant P and BM coupling strengths 
is not perfectly fulfilled. The entanglement entropy can therefore be utilized as an indirect quantitative measure 
of bosonic contamination of sought-after $W$ states due to excitation-boson coupling. Based on the results 
for the entanglement entropy obtained here [cf. Fig.~\ref{fig:EntropyPlot}] it can be inferred that this entropy 
shows a relatively weak growth as a function of the residual P coupling (originating from the nonvanishing 
difference between $g_{\textrm{P}}$ and $g_{\textrm{BM}}$) in the immediate vicinity of the ``sweet spot'' 
$g_{\textrm{P}}=g_{\textrm{BM}}$ of the model. This bodes well for the realization of $W$ states in 
either of the two proposed physical platforms.

\section{Summary and Conclusions} \label{SumConcl}
In summary, this paper investigated the ground-state entanglement spectrum and entropy of a model that describes the 
interplay of two of the most common mechanisms of short-ranged, nonlocal coupling of an itinerant spinless fermion excitation 
to zero-dimensional (dispersionless) bosons -- namely, the Peierls- and breathing-mode type interactions. In order
to be able to describe all the relevant physical regimes of this model, which displays sharp, level-crossing transitions 
at certain critical coupling strengths, the entanglement spectrum was evaluated in a numerically-exact fashion.  
The entanglement spectrum in the generic case of this model -- with unequal strengths of the two couplings -- is compared 
and contrasted with the special case of equal coupling strengths, in which the model supports bare-excitation Bloch eigenstates
(for an arbitrary coupling strength) and even a ground state of that same type below a critical coupling strength. 

It was demonstrated here that the behavior of the lowest entanglement-spectrum eigenvalue mimics that of the entanglement
entropy itself; most prominently, this eigenvalue vanishes -- like the entropy itself -- in the special case when the 
Peierls and breathing-mode couplings have the same strength. Moreover, it was shown that -- while reflecting the presence 
of sharp transitions through first-order nonanalyticities at several critical coupling strengths -- all the remaining 
entanglement-spectrum eigenvalues also show a singularity in the case of equal coupling strengths. Finally, it was 
demonstrated that the entanglement entropy shows only a weak growth in the vicinity of this ``sweet spot'' of the model.  
This behavior bodes well for the realization of multipartite $W$ states in superconducting and neutral-atom based qubit 
arrays that may serve as analog quantum simulators of the investigated model~\cite{Stojanovic:21,StojanovicPRL:20}; in 
those systems, bare-excitation ground states of this model translate into $W$ states. Furthermore, these same systems 
may also serve as platforms for an experimental measurement of the entanglement spectra computed in the present work 
using a previously proposed general method~\cite{Pichler+:16} based on an analogy to the many-body Ramsey 
interferometry~\cite{Ekert+:02}.

\begin{acknowledgments}
This research was supported by the Deutsche Forschungsgemeinschaft (DFG) -- SFB 1119 -- 236615297.
\end{acknowledgments}
\appendix 
\section{Derivation of the effective coupling strength} \label{derivelambda}
Here the expression for the effective e-ph coupling strength in the model at hand is obtained, by first finding 
the expressions for the P- and BM coupling strengths ($\lambda_{\textrm{P}}$ and $\lambda_{\textrm{BM}}$); the 
latter are derived starting from the general expression in Eq.~\eqref{genlambda}.

On account of the fact that the BM coupling depends only on $q$ [cf. Eq.~\eqref{gammaBM}], 
the Brillouin-zone average for this coupling can be expressed as 
\begin{equation}
\langle|\gamma_{\textrm{BM}}(q)|^{2}\rangle_{\textrm{BZ}}\equiv 
\frac{1}{2\pi}\:\int^{\pi}_{-\pi} |\gamma_{\textrm{BM}}(q)|^{2}\:dq \:,
\end{equation}
so that, as a special case of Eq.~\eqref{genlambda}, one arrives at
\begin{equation}
\lambda_{\textrm{BM}} = \frac{(2t_{\rm e}\:\omega_{\textrm{ph}})^{-1}}
{2\pi}\int^{\pi}_{-\pi}\:|\gamma_{\textrm{BM}}(q)|^{2}\:dq \:.
\end{equation}
By making use of Eq.~\eqref{gammaBM}, from the last equation one straightforwardly 
obtains
\begin{equation}
\lambda_{\textrm{BM}} = \frac{(2t_{\rm e}\:\omega_{\textrm{ph}})^{-1}}
{2\pi}\times 4g^2_{\textrm{BM}}\:\omega^2_{\textrm{ph}}\int^{\pi}_{-\pi} 
\sin^2 q\:dq \:, 
\end{equation}
which finally leads to
\begin{equation}
\lambda_{\textrm{BM}} = g^2_{\textrm{BM}}\:
\frac{\omega_{\textrm{ph}}}{t_{\rm e}}\:.
\end{equation}

At the same time, given that the P-coupling vertex function depends both on $k$ and $q$  [cf. Eq.~\eqref{gammaP}],
the corresponding Brillouin-zone average ought to involve integrations over both excitation- and phonon 
quasimomenta, i.e., 
\begin{equation}
\langle|\gamma_{\textrm{P}}(k,q)|^{2}\rangle_{\textrm{BZ}}\equiv\frac{1}{(2\pi)^2}
\:\int^{\pi}_{-\pi}\int^{\pi}_{-\pi}\:|\gamma_{\textrm{P}}(k,q)|^{2}\:dkdq  \:.
\end{equation}
Therefore, $\lambda_{\textrm{P}}$ is given by [cf. Eq.~\eqref{genlambda}]
\begin{equation}
\lambda_{\textrm{P}} = \frac{(2t_{\rm e}\:\omega_{\textrm{ph}})^{-1}}{(2\pi)^2} 
\int^{\pi}_{-\pi}\int^{\pi}_{-\pi}\:|\gamma_{\textrm{P}}(k,q)|^{2}\:dkdq \:,
\end{equation}
which, using Eq.~\eqref{gammaP}, further reduces to
\begin{eqnarray}\label{pretpeq}
\lambda_{\textrm{P}} &=& \frac{(2t_{\rm e}\:\omega_{\textrm{ph}})^{-1}}
{(2\pi)^2}\:\times 4g^2_{\textrm{P}}\:\omega^2_{\textrm{ph}} \nonumber \\
&\times& \int^{\pi}_{-\pi}\int^{\pi}_{-\pi} [\sin(k+q)-\sin k]^2\:dkdq \:.
\end{eqnarray}
By evaluating the integral in the last equation,
\begin{equation}
\int^{\pi}_{-\pi}\int^{\pi}_{-\pi} [\sin(k+q)-\sin k]^2\:dkdq = 4\pi^2 \:,
\end{equation}
and inserting this last result into Eq.~\eqref{pretpeq}, one finally obtains 
\begin{equation}
\lambda_{\textrm{P}} = 2g^2_{\textrm{P}}\:\frac{\omega_{\textrm{ph}}}
{t_{\rm e}}\:.
\end{equation}

It is worthwhile noting that the total effective e-ph coupling strength $\lambda_{\textrm{e-ph}}$ 
[cf. Eq.~\eqref{genlambda}] is here given by the simple sum of $\lambda_{\textrm{P}}$ and 
$\lambda_{\textrm{BM}}$, i.e. 
\begin{equation}
\lambda_{\textrm{e-ph}}=\lambda_{\textrm{P}}+\lambda_{\textrm{BM}}
= (2g^2_{\textrm{P}}+g^2_{\textrm{BM}})\:\frac{\omega_{\textrm{ph}}}{t_{\rm e}} \:.
\end{equation}
Namely, while the expression for $\langle|\gamma_{\textrm{e-ph}}(k,q)|^{2}\rangle_{\textrm{BZ}}$
[cf. Eq.~\eqref{genlambda}] in the case of simultaneous P and BM couplings also contains
the cross terms originating from the product of the P and BM vertex functions [cf. 
Eqs.~\eqref{gammaP} and \eqref{gammaBM}], it is straightforward to show that their BZ 
average -- which entails integrals over $k,q\in (-\pi, \pi]$ -- is equal to zero; thus, 
those terms do not contribute to $\lambda_{\textrm{e-ph}}$.

\section{Derivation of the reduced density matrix $\rho^{(\textrm{gs})}_{\textrm{e}}$} \label{RedDensMatEl}
In what follows, an explicit derivation is provided of the expression for the matrix elements 
of the reduced density matrix $\rho^{(\textrm{gs})}_{\textrm{e}}$ [cf. Eq.~\eqref{rhoematrel}] corresponding to 
the ground state $|\psi_{\textrm{gs}}\rangle$ of the system. 

The ground state $|\psi_{\textrm{gs}}\rangle$ belongs to the $K=K_{\textrm{gs}}$ sector of 
the Hilbert space of the system [recall the discussion in Sec.~\ref{CompStrategy}].
Therefore, this state can be expanded as [cf. Eq.~\eqref{decomp_psigs}]
\begin{equation}\label{decomp_Kgs}
|\psi_{\textrm{gs}}\rangle=\sum_{\mathbf{m}}C_{K_{\textrm{gs}},\mathbf{m}}
|K_{\textrm{gs}},\mathbf{m}\rangle \:,
\end{equation}
where $|K_{\textrm{gs}},\mathbf{m}\rangle$ is the symmetry-adapted basis of that 
Hilbert-space sector [cf. Eq.~\eqref{symmbasalter}]:
\begin{equation}\label{symmbas_Kgs}
|K_{\textrm{gs}},\mathbf{m}\rangle = N^{-1/2} \sum_{n=1}^N e^{iK_{\textrm{gs}} n}
\,|n\rangle_\text{e}\otimes \mathcal{T}^{\textrm{ph}}_{n-1}|\mathbf{m}\rangle_\text{ph} \:.
\end{equation}
By making use of the expansion in Eq.~\eqref{decomp_Kgs}, the corresponding density matrix 
of the e-ph system can be written in the form
\begin{equation} \label{rhoeph}
\rho^{\textrm{(gs)}}_{\textrm{e-ph}}=\sum_{\mathbf{m},\mathbf{m'}}\:C^{*}_{K_{\textrm{gs}},\mathbf{m'}}
\:C_{K_{\textrm{gs}},\mathbf{m}}\:|K_{\textrm{gs}},\mathbf{m}\rangle\langle K_{\textrm{gs}},
\mathbf{m'}|\:.
\end{equation}
Using Eq.~\eqref{symmbas_Kgs}, it further follows that
\begin{eqnarray}\label{rhoeph}
\rho^{\textrm{(gs)}}_{\textrm{e-ph}}&=& N^{-1}\sum_{\mathbf{m},\mathbf{m'}}\sum^{N}_{n,n'=1}\:
e^{iK_{\textrm{gs}} (n-n')}C^{*}_{K_{\textrm{gs}},\mathbf{m'}}\:C_{K_{\textrm{gs}},\mathbf{m}}\: \nonumber\\
&\times&|n\rangle\langle n'|\otimes|\mathcal{T}^{\textrm{ph}}_{n-1}\mathbf{m}\rangle
\langle \mathcal{T}^{\textrm{ph}}_{n'-1}\mathbf{m'}|\:.
\end{eqnarray}

According to Eq.~\eqref{rho_e}, the reduced excitation density matrix $\rho^{(\textrm{gs})}_{\textrm{e}}$ can now be 
obtained by tracing the density matrix $\rho^{\textrm{(gs)}}_{\textrm{e-ph}}$ over the phonon basis. In other words,
\begin{equation}
\rho^{(\textrm{gs})}_{\textrm{e}}=\sum_{\mathbf{m''}}\langle \mathbf{m''}|\:
\rho^{\textrm{(gs)}}_{\textrm{e-ph}}\:|\mathbf{m''}\rangle\:,
\end{equation}
where $\mathbf{m''}$ is the dummy index for the phonon basis states [i.e., $\{\mathbf{m''}\}$ represents 
the set of all phonon occupation-number configurations]. By inserting $\rho^{\textrm{(gs)}}_{\textrm{e-ph}}$ from Eq.~\eqref{rhoeph}, 
one obtains 
\begin{eqnarray}\label{exprrhoe}
\rho^{(\textrm{gs})}_{\textrm{e}}&=&N^{-1}\sum_{\mathbf{m},\mathbf{m'},\mathbf{m''}}
\sum^{N}_{n,n'=1}\:e^{iK_{\textrm{gs}}(n-n')}C^{*}_{K_{\textrm{gs}},\mathbf{m'}}
\:C_{K_{\textrm{gs}},\mathbf{m}}\:\nonumber\\
&\times& \:\langle\mathcal{T}^{\textrm{ph}}_{n'-1}\mathbf{m'}|\mathbf{m''}\rangle   
\langle \mathbf{m''}|\mathcal{T}^{\textrm{ph}}_{n-1}\mathbf{m}\rangle\:
|n\rangle\langle n'|\:.
\end{eqnarray}

At this point it is convenient to first carry out the summation over $\mathbf{m''}$. 
By making use of the completeness relation in the phonon Hilbert space
\begin{equation}\label{}
\sum_{\mathbf{m''}}|\mathbf{m''}\rangle \langle \mathbf{m''}|
=\mathbbm{1}_{D_{\textrm{ph}}}\:,
\end{equation}
where $\mathbbm{1}_{D_{\textrm{ph}}}$ is the identity operator in that space, it is 
straightforward to verify that
\begin{equation}\label{interm}
\sum_{\mathbf{m''}}\langle\mathcal{T}^{\textrm{ph}}_{n'}\mathbf{m'}|\mathbf{m''}\rangle   
\langle \mathbf{m''}|\mathcal{T}^{\textrm{ph}}_{n}\mathbf{m}\rangle=
\langle\mathcal{T}^{\textrm{ph}}_{n'-1}\mathbf{m'}|\mathcal{T}^{\textrm{ph}}_{n-1}
\mathbf{m}\rangle \:.
\end{equation}
Using this last result, the expression for $\rho^{(\textrm{gs})}_{\textrm{e}}$ in Eq.~\eqref{exprrhoe} 
now reduces to
\begin{eqnarray}\label{rhoefinal}
\rho^{(\textrm{gs})}_{\textrm{e}} &=& N^{-1}\sum_{\mathbf{m},\mathbf{m'}}\sum^{N}_{n,n'=1}\:
e^{iK_{\textrm{gs}}(n-n')}C^{*}_{K_{\textrm{gs}},\mathbf{m'}}\:C_{K_{\textrm{gs}},\mathbf{m}}\:\nonumber\\
&\times& \:\langle\mathcal{T}^{\textrm{ph}}_{n'-1}\mathbf{m'}|\mathcal{T}^{\textrm{ph}}_{n-1}
\mathbf{m}\rangle\:|n\rangle\langle n'|\:.
\end{eqnarray}
The last expression implies that the matrix elements of the reduced excitation density 
matrix are given by
\begin{eqnarray}\label{rhoematrel}
\left(\rho^{(\textrm{gs})}_{\textrm{e}}\right)_{nn'}&=& N^{-1}\:e^{iK_{\textrm{gs}}(n-n')}\sum_{\mathbf{m},\mathbf{m'}}
\:C^{*}_{K_{\textrm{gs}},\mathbf{m'}}\:C_{K_{\textrm{gs}},\mathbf{m}}\nonumber \\
&\times&\:\langle\mathbf{m'}\:|\:\mathcal{T}^{\textrm{ph}}_{n-n'}\mathbf{m}\rangle\:,
\end{eqnarray}
where use has been made of the fact that $\langle\mathcal{T}^{\textrm{ph}}_{n'}\mathbf{m'}\:|\:
\mathcal{T}^{\textrm{ph}}_{n}\mathbf{m}\rangle\equiv \langle\mathbf{m'}\:|\:
\mathcal{T}^{\textrm{ph}}_{n-n'}\mathbf{m}\rangle$.

\end{document}